# Destruction of the Mott Insulating Ground State of $Ca_2RuO_4$ by a Structural Transition

C. S. Alexander[1], G. Cao[1*], V. Dobrosavljevic[1], E. Lochner[2], S. McCall[1], J. E. Crow[1] and R. P. Guertin[3]

[1]National High Magnetic Field Laboratory, 1800 East Paul Dirac Drive, Tallahassee, FL 32310
[2]Center for Material Research and Technology, Florida State University, Tallahassee, FL 32306
[3]Department of Physics and Astronomy, Tufts University, Medford, MA 02155

We report a first-order phase transition at $T_M$=357 K in single crystal $Ca_2RuO_4$, an isomorph to the superconductor $Sr_2RuO_4$. The discontinuous decrease in electrical resistivity signals the near destruction of the Mott insulating phase and is triggered by a structural transition from the low temperature orthorhombic to a high temperature tetragonal phase. The magnetic susceptibility, which is temperature dependent but not Curie-like decreases abruptly at $T_M$ and becomes less temperature dependent. Unlike most insulator to metal transitions, the system is not magnetically ordered in either phase, though the Mott insulator phase is antiferromagnetic below $T_N$=110 K.

The d-shell orbitals of the 4d and 5d transition metal cations in transition metal oxides (TMO) are more extended than those of their more thoroughly studied counterparts, the 3d TMO's. On the one hand, the extended d-shells suggest *a priori* a weaker intra-atomic Coulomb interaction, U, relative to the 3d's. Within the context of the Mott-Hubbard model this alone would suggest metallic behavior, with U/W<1, W being the bandwidth, which is proportional to the near neighbor hopping probability. On the other hand, the extended d-shells suggest a robust interaction betweem d-orbitals and nearest neighbor oxygen orbitals, and as a result some 4d- and 5d-TMO's have a tendency to form structures which are distortions of the more ideal generic structure, such as the $K_2NiF_4$ structure on which the subject of this report is based. In these cases the M-O-M bond angle may be considerably less than the ideal 180º and this generally implies insulating behavior because of a narrowing of the d-electron bandwidth[1]. Given the competition between these two effects, it is not surprising that some 4d and 5d TMO's verge on the interface between metal and insulator with U/W ≈1. Owing to such a sensitive U/W ratio, small perturbations, such as slight alterations of crystal structure, dilute doping, etc., can readily tip the balance across the metal-nonmetal boundary, with resultant pronounced changes in physical properties. This feature has been well-illustrated in recent studies on layered ruthenates and iridates[2-4].

In this paper, we report a strongly first order phase transition at ambient pressure from a low temperature orthorhombic structure Mott insulating phase to a high temperature tetragonal "near" metallic phase at $T_M$=357 K in stoichiometric single crystal $Ca_2RuO_4$. Much of the interest in $Ca_2RuO_4$ derives from it being isostructural with $Sr_2RuO_4$, the only non-cuprate superconductor with a high $T_c$ cuprate structure and which likely has "p-wave" rather than conventional pair coupling[5]. In contrast to spontaneous metal-insulator transitions which normally accompany a transition from low temperature antiferromagnetism to paramagnetism (e.g., $T_M$=150 K in $V_2O_3$[6] and $T_M$=48 K in $Ca_3Ru_2O_7$[2]), $Ca_2RuO_4$ is paramagnetic on both sides of the transition, with well defined magnetic moments associated with the Ru ions. (Ordered antiferromagnetic coupling exists only below $T_N$= 110 K.) The elastic and electronic degrees of freedom are closely coupled in this system. Therefore, we propose that the origin of the transition is an energy balancing tradeoff between the saving of elastic energy as the system transforms from a state of low to high symmetry at the expense of electronic energies from a phase of high resistance, low entropy to a phase of lower resistance, higher entropy. We further propose that the transition at T=357 K which destroys the Mott insulating phase qualifies as a Mott transition, based on Mott's own generalization of the definition of such a transition [Ref. 7, pp 138-9].

The results of x-ray diffraction (lattice parameters), electrical resistivity and magnetization are presented as a function of temperature for 70<T<600 K in single crystal $Ca_2RuO_4$, which is a recently discovered compound[3,8] belonging to the layered Ruddlesden-Popper series with a single Ru-O layer in the unit cell. Our motivation for this study extends from prior work where the resistivity of $Ca_2RuO_4$ was found to decrease by an astonishing eight orders of magnitude (from $10^{10}$ to 100 Ω-cm between 70<T<300 K)[3]. This suggested a rapid temperature-driven decrease in U/W, which could eventually lead to an insulator-to-metal transition at a more elevated temperature. The transition is also predicted by a simple extrapolation of the temperature dependent neutron scattering studies of Braden et al[9].

There have been a handful of reported studies on $Ca_2RuO_4$ to date. Based on these studies, a few major features are known[3,8,9]: (1) It has the $K_2NiF_4$ structure with large rotations and tilts of the $RuO_6$ octahedra compared to $Sr_2RuO_4$ and due to the smaller ionic radius of Ca; (2) it appears to be a Mott insulator below T=300 K with a narrow gap of about 0.2 eV, as determined from both transport and optical measurements[10]; (3) it is antiferromagnetically ordered below $T_N$=110 K; and (4) the magnetic moment (≈0.4 $\mu_B$/Ru at 30 T) is low compared to the other Ca- and Sr-ruthenates[3] which have moments closer to the 2.0 $\mu_B$/Ru expected for the low spin (S=1) state of the $Ru^{4+}(4d^4)$ configuration.

Single crystals of $Ca_2RuO_4$ were grown in Pt crucibles using a flux technique described elsewhere[2]. X-ray diffraction for 90<T<400 K was performed on powdered single crystals using Siemens Θ-2Θ and Θ-Θ diffractometers with low and high temperature attachments. The metric refinement was carried out using 21 reflections. Resistivity was measured with a standard four probe technique and magnetization with a commercial SQUID magnetometer. All results of x-ray difffraction and EDX indicate that the crystals studied are pure without any second phase.

Shown in Fig. 1a is the temperature dependence of the lattice parameters for 90<T<400 K revealing a sharp transition near $T_M$=357 K from a low temperature orthorhombic phase to a high temperature tetragonal phase. The results for T<300 K agree reasonably well with those described in Ref. 9, which were derived from neutron measurements on polycrystalline $Ca_2RuO_4$ for 11<T<300 K. This phase transition is well characterized by splittings of ($l$00) or (0k0) peaks in the temperature-dependent diffraction patterns. Below $T_M$, the a-axis decreases whereas the b-axis increases. As temperature decreases over the interval from 400 K to 90 K, the a-axis contracts by 1.5% and the b-axis expands by 3%. The positive and negative thermal expansion coefficients derived from the temperature dependence of the lattice parameters not only indicate an increasingly strong orthorhombic distortion in the Ru-O plane but also are conspicuously large, i.e., one order of magnitude larger than those for other related compounds such as $Sr_2IrO_4$ [11], and $SrRuO_3$[12], $CaRuO_3$[12] and $Sr_2RuO_4$[13], which do not undergo first order phase transitions. The lattice volume is also substantially changed by 1.3% (see Fig. 1b). It is therefore not surprising that such a drastic structural change is even macroscopically visible: The crystals shatter when they were heated through the transition temperature. (The shattering makes the resistivity measurements a difficult task: The measurements had to be performed on extremely small residual pieces of shattered crystals which are approximately 0.3x0.3x0.1 $mm^3$).

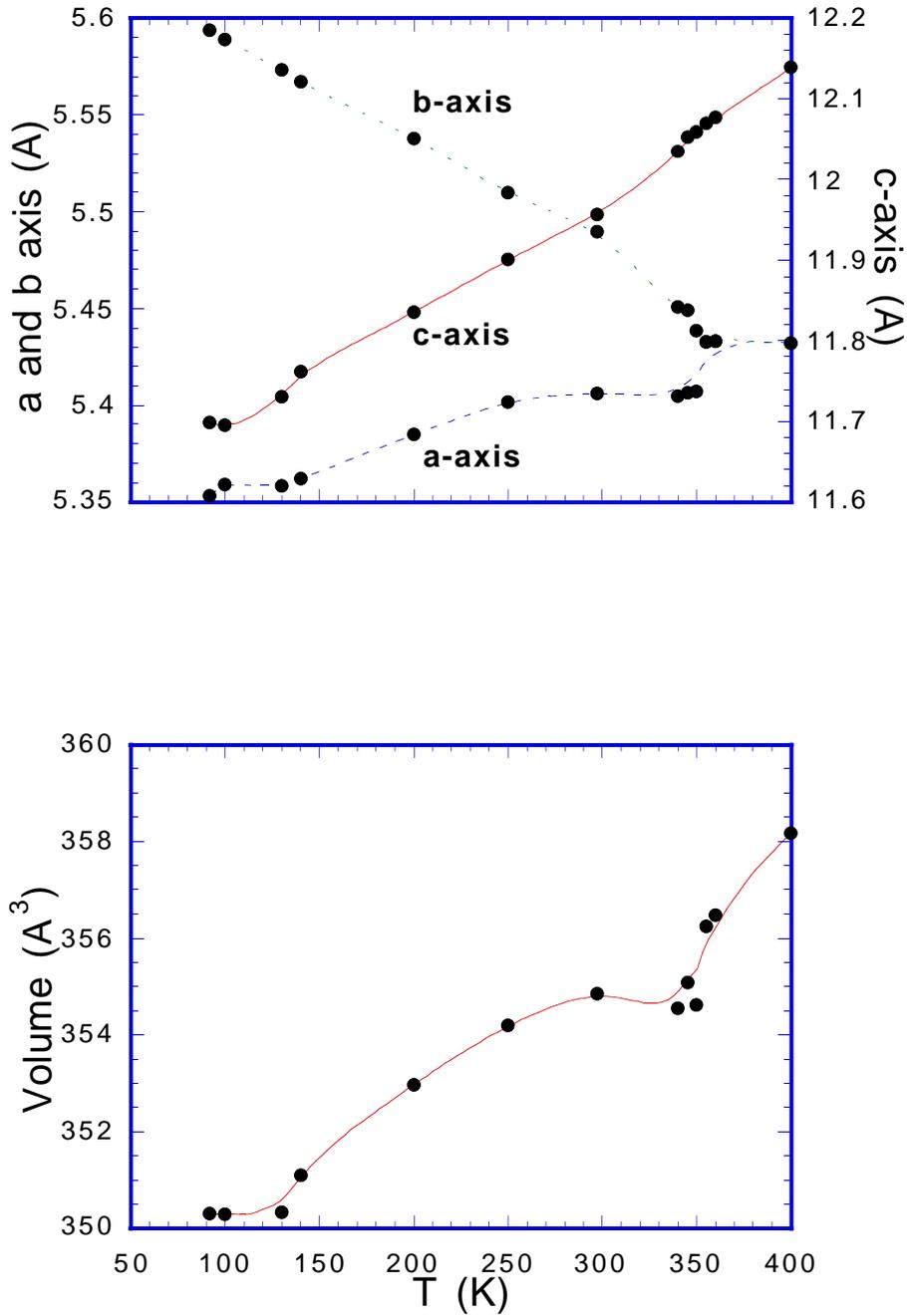

Fig. 1. Temperature dependence of lattice parameters for 90<T<400 K for the powdered single crystal $Ca_2RuO_4$.

Although the sensitivity of our powder x-ray diffraction data is not sufficient to obtain a complete structural determination, the phase transition at T=357 K is most likely caused by a rotation and tilt of the $RuO_6$ octahedra. According to the results in Ref. 9, the presence of the orthorhombic distortion below T=300 K is due to a combination of a rotation of the $RuO_6$ octahedra around the c-axis (11.8°) and a temperature-dependent tilt of the Ru-O basal planes (11.2-12.7°). A similar structural phase transition due to basal plane tilting is also observed in isomorphic compounds $La_2NiO_4$ and $La_2CuO_4$ although the orthorhombic distortion is much less severe[14] and there is no corresponding impact on the electrical conductivity.

Fig. 2 shows electrical resistivity, $\rho(T)$, in the ab-plane as a function of temperature for 70<T<600 K. (The resistivity results up to room temperature as well as the those of the magnetic properties for 10<T<300 K agree reasonably well with those of Ref. 8 for the "S" phase.) An abrupt transition from an insulating state to a nearly metallic state occurs at $T_M$=357 K, simultaneous with the structural transition. The sudden decrease in $\rho$ by a factor of 3.5 at the transition observed in several well-characterized crystals results in a pronounced discontinuity (see inset) indicating a robust first-order transition and unambiguously characterizes a discontinuous alteration in the d-band structure typical of a metal-insulator transition. Results of a recent optical conductivity study on $Ca_2RuO_4$ give an optical gap of 0.2 eV at T=300 K[10], consistent with the lack of antiferromagnetism in that temperature region [Ref. 7, p. 176]. Below the transition, $\rho(T)$ rises rapidly, increasing eight orders of magnitude over a relatively narrow temperature interval. More remarkably, $\rho(T)$ can be well fit for 70<T<300 K to variable-range hopping or the Efros-Shklovskii mechanism given by $\rho(T)=A\exp(T_o/T)^\nu$ with $\nu=1/2$. We note a discontinuity in $d\rho/dT$ at T≈250 K (though not in $\rho(T)$) in the c-axis resistivity of the iso-structural system $(La_{1-x}Sr_x)_2CuO_4$ for x≈0.10, presumably due to a change in incoherent hopping[15].

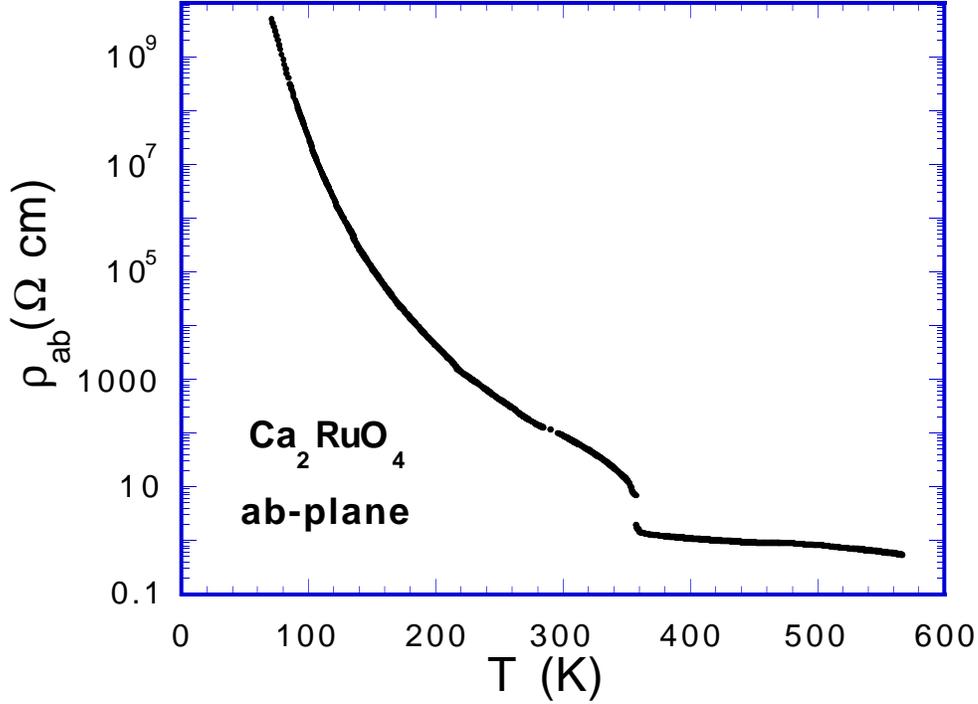

Fig. 2. Electrical resistivity $\rho(T)$ for the ab-plane as a function of temperature for $70<T<600$ K.

We note that the metallic state is not fully realized above the transition. For this system, the "near" metallic behavior may, in fact, be an artifact of the required method of measurement, i.e., the temperature must be <u>increased</u> from room temperature through $T_M$, and some microcracking of the sample occurs as evidenced by the inability to reproduce all the data upon lowering the temperature. In lightly La-doped samples, $Ca_{1-x}La_xRuO_4$ for $x<0.05$ (unpublished data), where $T_M$ is below room temperature, all samples show clear <u>metallic</u> behavior as temperature is <u>reduced</u> through $T_M$, which remains a first order transition to an insulating phase. Then upon raising the temperature through $T_M$, "near" metallic behavior is observed. In any case, such behavior is not limited to this system. "Near" metallic behavior is seen in $(V_{1-x}Cr_x)_2O_3$, the most widely cited metal insulator transition system in the literature, for Cr concentrations very close to that which completely suppresses the metallic phase (Ref. 7, Fig. 6.3).

Although due to different causes, ρ above the Verwey transition at T=120 K in $Fe_3O_4$ is surprisingly similar to that observed in Fig. 2[7,16]. (Several examples of near metallic conductivity above $T_M$ in perovskite related systems, including $La_{1-x}Sr_xFeO_3$, are found in Ref. 17, Section IV.)

Fig. 3 shows the magnetic susceptibility defined as M/H vs. temperature for the field along the ab-plane, featuring the antiferromagnetic transition at $T_N$=110 K[3,8]. At T=357 K there is a small yet well defined anomaly. While the shape might suggest the onset of ferromagnetism, there is neither hysteresis nor negative curvature of the isothermal magnetization normally expected for a ferromagnet for T<357 K (see inset to Fig. 3), so the system remains paramagnetic through the structural/metal insulator transition.

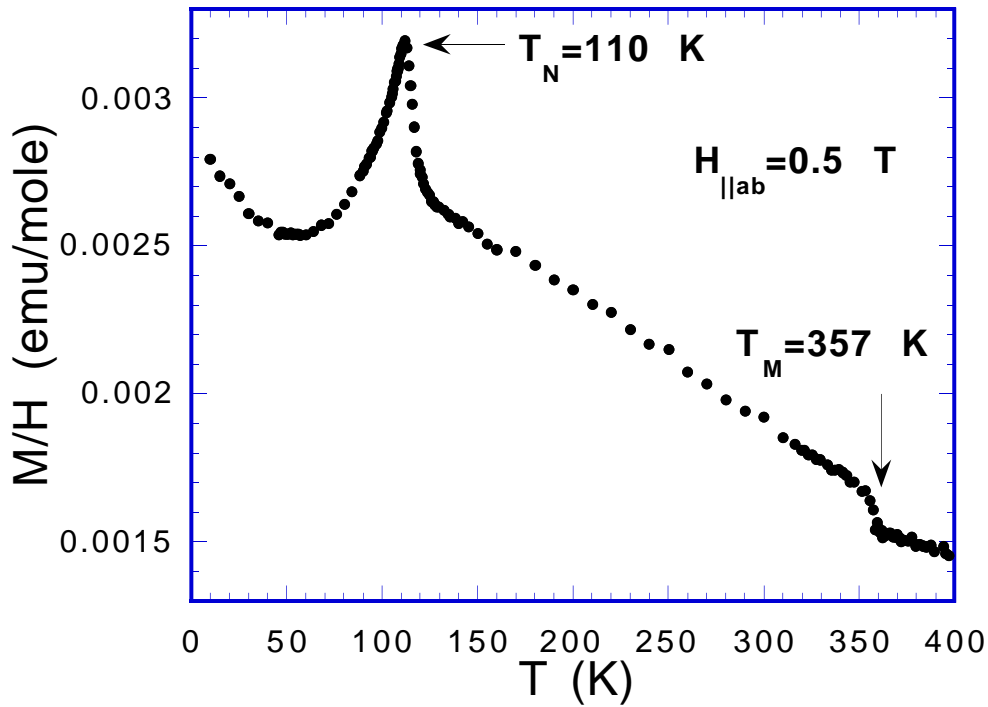

Fig. 3. Magnetic susceptibility defined as M/H for the ab-plane as a function of temperature for 2<T<400 K

It is puzzling that M/H vs T is effectively "negative linear" for 130<T<350 K, clearly not Curie-Weiss (C-W)-like. Nakatsuji et al observed a similar effect [8]. The sample volume changes extensively (+1.3%-see Fig. 1b) over this temperature interval. However, a simple increase in volume should reduce $\chi(T)$ slightly, retaining C-W behavior. (Early work on U(Fe) alloys [18] tried to take thermal expansion into account to explain directly $\chi(T)$ deviations but was not successful.) We postulate, therefore, that the unstable lattice structure and the concommitant change in band structure is indirectly responsible for the anomalous behavior: The magnetic moment at individual sites is probably a strong function of bandwidth (hopping probability). This instability drives a strong temperature dependence of the M-O-M bond angle, which controls the bandwidth and indirectly the individual site moments. This is somewhat analogous to the case for a-Y(Fe) alloys[19] where temperature dependent spin fluctuations link anomalous $\chi(T)$ with large volume changes. It is significant that the effect of thermal expansion on $\chi(T)$ in a-Y(Fe) occurred also in the non-magnetically ordered region of temperature.

The sharp peak in M/H at $T_N$=110 K is attributed to an antiferromagnetic transition, but no corresponding change in $\rho$ is discerned[3] . This is understandable because for an antiferromagnetic insulator the magnitude of the Hubbard gap is not greatly affected through $T_N$ (Ref. 7, p. 137). The spin structure associated with this Neel phase is quite sensitive to impurity doping: 1% La doping for Ca, for instance, can effectively alter the spin configuration from antiferromagnetic to apparent ferromagnetic coupling, with a Curie temperature $T_C$=135 K. The sensitivity of the spin configuration to the impurity doping reflects the subtlety of the magnetic structure which, again may be related to the instability of the distorted crystal structure.

While it cannot be totally ruled out that the abrupt change in resistivity at $T_M$ merely reflects a massive change in incoherent hopping conductivity at a structural transition and hence not a true metal-insulator transition, the confluence of structural, transport and magnetic anomalies at this first order transition, coupled with the anomalous magnetization for $T<T_M$ suggests otherwise. In $Ca_2RuO_4$ the low temperature phase, $T_N<T<T_M$, is characterized by a relatively small electronic gap in N(E) ($\approx$0.2 eV) due to the rotation and tilt of the $RuO_6$ octahedra, which is known to minimize the total energy of the system. If the gain in elastic energy, $E_E$, is larger than the loss of the electronic coherence energy $E_C$, then the system will be a paramagnetic Mott insulator at low temperatures. On the other hand, one expects the $E_E$ gain to be reduced as the temperature is increased, and the associated rotation or tilt angle $\Theta$ will decrease. As a result, the electronic bandwidth $W \approx \cos(\Theta/2)$ will increase to the point that it will be more favorable for the system to transform to a metallic (correlated Fermi liquid) state and gain a coherence energy $E_C$. Thus, the thermally-induced vibrations will reduce the overall distortion of the $RuO_6$ octahedra, and ultimately increase the electronic bandwidth and stabilize a paramagnetic metallic state at high temperatures. This structurally-driven Mott transition can be straightforwardly modeled by extending the existing models to incorporate the relevant coupling to elastic degrees of freedom. Preliminary calculations indicate that the expected PMI to PMM transition can follow, the detailed results will be presented elsewhere.

We emphasize that the transition at $T_M=357$ K is not driven or even indirectly influenced by a magnetic ordering instability, which in the case of antiferromagnetism and the associated band splitting may help stabilize a low temperature insulator, as recognized early by Slater. Of course, at the lower temperatures one does expect exchange coupling between the localized magnetic moments in the insulator which will lead to an antiferromagnetically ordered state, and indeed such a Neel phase is observed below $T_N=110$ K.
The opposite type of temperature-driven metal-insulator transition from high temperature paramagnetic insulator (PMI) to a lower temperature paramagnetic metal (PMM) has been experimentally observed in $(V_{1-x}Cr_x)_2O_3$ for $x=0.51$[6,7]. The lower temperature metallic state is viewed as a highly correlated Fermi liquid characterized by a large effective mass m* and a low coherence temperature $T^* \approx (m^*)^{-1}$. At temperatures $T>T^*$, a transition to a paramagnetic insulating state is predicted, where electrons become localized magnetic moments. This entropy gain overwhelms the energy loss $E_C \approx k_B T^*$ due to the destruction of the coherent Fermi liquid and serves as a driving force for this transition.

The dynamical mean field theory describing the above transition requires an electronic system with a fixed bandwidth, W

and an on-site Coulomb repulsion U which is a function of temperature. In real systems the W can also be a strong function of temperature as described above, so that the proposed theoretical picture has to be modified. Consequently, in $Ca_2RuO_4$, where the sequence of phases is reversed, strong temperature dependence to the bandwidth may drive the transition. Evidence that the bandwidth has such temperature dependence comes from two sources: First, the M-O-M bond angle is a strong controller of bandwidth[1] - the smaller the angle, the narrower the band. For $Ca_2RuO_4$, the M-O-M bond angle at T=11 K is only 151°, far from the ideal 180°. This angle increases with increasing temperature, as evidenced in Fig. 1, driving the system towards a more metallic phase. Second, we expect the localized paramagnetic moments to be a function of bond angle, and the anomalous (non-Curie) behavior for $T_N<T<T_M$ reflects this effect (see Fig. 3).

In conclusion, a structurally-driven transition from a nearly metallic state to an insulating state is observed at T=357 K. We have argued that the transition is Mott-like and not associated with a magnetic transition, and it is driven primarily by the coupled electronic elastic energies associated with the structural change. Generally, in a Mott insulator the gap depends only on the existence of moments and is not coupled to the crystal structure[7]. In contrast the simultaneous structural and electronic transition at T=357 K in $Ca_2RuO_4$ clearly points out that the transport properties are chiefly driven by changes in elastic energy. Though this transition is the most dramatic manifestation of the control exerted by the crystalline phase on the transport properties through bandwidth, there is evidence for this control throughout the temperature range covered.

The authors at NHMFL wish to acknowledge support provided by the National Science Foundation under Cooperative Agreement No. DMR95-27035 and the State of Florida. V.D. was partially supported by the Alfred P. Sloan Foundation. R.P.G. was partially supported by the Research Corporation.

tend to saturate for T<120 K. This is similar to the invar effect observed below $T_C$=165 K in $SrRuO_3$ by these authors.